\documentclass{vgtc}                          
\ifpdf
  \pdfoutput=1\relax                   
  \usepackage{graphicx}                
  \DeclareGraphicsExtensions{.pdf,.png,.jpg,.jpeg} 
\else
  \usepackage{graphicx}                
  \DeclareGraphicsExtensions{.eps}     
\fi%

\graphicspath{{figures/}{pictures/}{images/}{./}} 
\usepackage[dvipsnames]{xcolor}
\usepackage{microtype}                 
\PassOptionsToPackage{warn}{textcomp}  
\usepackage{textcomp}                  
\usepackage{mathptmx}                  
\usepackage{times}                     
\usepackage{cite}                      
\usepackage{tabu}                      
\usepackage{booktabs}                  
\usepackage{enumitem}                  
\usepackage{tabu}                      
\usepackage{booktabs}                  
\usepackage{algorithm}
\usepackage{algorithmic}

\usepackage[normalem]{ulem}

\definecolor{codeGreen}{rgb}{0,0.6,0}

\newcommand{\edit}[1]{{\color{black}#1}}
\newcommand{\second}[1]{{\color{black}#1}}

\newcommand{\dataset}[1]{``#1''}
\newcommand{\systemname}{\emph{LabelVizier}}
\newcommand{\viewA}{Label Investigation View}
\newcommand{\viewB}{Record Projection View}
\newcommand{\viewC}{Inspection \& Operation View}

\onlineid{8193}

\vgtccategory{Technique}

\vgtcinsertpkg

\title{ \systemname: Interactive Validation and Relabeling for \\Technical Text Annotations}

\author{Xiaoyu Zhang\thanks{Equally contributed.} \thanks{e-mail: \{xybzhang, xwxuan, klma\}@ucdavis.edu}\\ %
        \scriptsize University of California, Davis %
\and Xiwei Xuan\footnotemark[1] \footnotemark[2]\\ %
     \scriptsize University of California, Davis %
\and Alden Dima\thanks{e-mail: \{alden.dima, thurston.sexton\}@nist.gov}\\ %
    \scriptsize National Institute of Standards and Technology
\and Thurston Sexton\footnotemark[3]\\ %
    \scriptsize National Institute of Standards and Technology
\and Kwan-Liu Ma\footnotemark[1]\\ %
     \scriptsize University of California, Davis
}

\shortauthortitle{Zhang \MakeLowercase{\textit{et al.}}: \systemname: Interactive Validation and Relabeling for Technical Text Annotations}

\abstract{
With the rapid accumulation of text data produced by data-driven techniques, the task of extracting ``data annotations''---concise, high-quality data summaries from unstructured raw text---has become increasingly important.
The recent advances in weak supervision and crowd-sourcing techniques provide promising solutions to efficiently create annotations (labels) for large-scale technical text data.
However, such annotations may fail in practice because of the change in annotation requirements, application scenarios, and modeling goals, where label validation and relabeling by domain experts are required. 
To approach this issue, we present \systemname{}, a human-in-the-loop workflow that incorporates domain knowledge and user-specific requirements to reveal actionable insights into annotation flaws, then produce better-quality labels for large-scale multi-label datasets.
We implement our workflow as an interactive notebook 
to facilitate flexible error profiling, in-depth annotation validation for three error types, and efficient annotation relabeling on different data scales. 
We evaluated
our workflow
\edit{in assisting the validation and relabelling of technical text annotation}
with two use cases and four expert reviews.
The results show that \systemname{} is applicable in various application scenarios, 
\edit{and users with different knowledge backgrounds have diverse preferences for the tool usage. }
}


\begin{document}



\firstsection{Introduction}

\maketitle

\vspace{-0.5mm}
Data-driven approaches have pervaded 
manufacturing in the age of Industry 4.0, producing a large amount of digitized data in the form of unstructured technical text~\cite{risch2008text}.
\edit{For example in machine maintenance, machine operators and repairing technicians often create maintenance work orders (MWOs) to record their maintenance activities.}
However, the rich text of asset management history in MWOs usually sits untouched because of the potential inconsistency, incompleteness, or incorrectness~\cite{brundage2021technical} in the descriptive text.
Compared to raw unstructured text, a set of high-quality annotations summarizing the content is \edit{preferred} for ``robust and reproducible''~\cite{brundage2021technical} analysis of large-scale technical text.
In particular, these annotations can be utilized for the systematic problem identification and classification, root cause analysis, and product life cycle prediction~\cite{fort2016collaborative}, which provides precious insights and facilitate the key performance index (KPI) assessment and budget planning process.
For instance, the statistics of the label ``\emph{too\_hot}'' in a heating, ventilation, and air conditioning (HVAC) system maintenance log dataset (see Sec.~\ref{subsec:use_case_HVAC}) could indicate how well the air conditioning system has been maintained and thus inform maintenance budget planning.
This is also a critical research topic in technical language processing (TLP)~\cite{Dima2021-ba}.
However, it is not easy to create quality annotations and many important annotated datasets are riddled with labelling errors~\cite{Northcutt2021-kl}.
Given the exponentially increasing volume of unstructured text, 
researchers have gradually discarded conventional manual annotation approaches and turned to more efficient state-of-the-art machine learning (ML) techniques or commercial crowd-sourcing~\cite{jeff2006crowdsource} platforms.
Particularly, recent advances in weak supervision~\cite{ratner2017snorkel, varma2018snuba} promise efficient large-scale text annotation.
However, it is necessary to sufficiently validate and improve the annotations generated by such methods before delivering them to down-streaming tasks.
Limited research efforts have been devoted \edit{to} validation and relabeling of such large-scale technical text annotation.
To facilitate this process, we developed \systemname{}, a human-in-the-loop workflow encapsulated as a visual analytic solution that supports reliable and efficient annotation validation and relabelling for domain practitioners to meet their specific application requirements.
\systemname{} helps identify and correct three types of annotations errors: (1) duplicate, (2) wrong, and (3) missing labels.
Inspired by the practice of debugging in software engineering, we profile the potential errors in the existing labels and devise visual analytics procedures to facilitate an efficient skimming of the labels and their context based on the domain expert's annotation preferences.
We supplement this validation process by training a surrogate model to approximate the agnostic annotation process, visualizing the prediction metadata to expose potential errors, and providing LIME explanations~\cite{ribeiro2016should} for root cause analysis.
For the user-identified annotation errors, we support flexible relabelling of the dataset on the corpus, sub-group, and record levels.
We implement this workflow as a web-based interactive notebook containing editable function blocks 
and an interactive visual analytic interface designed in close collaboration with the two domain experts on our team.
We demonstrate how \systemname{} can benefit different application scenarios in two use cases and evaluate them with expert reviews from four domain practitioners.
The results show that the domain experts appreciated the efficiency and accessibility of \systemname{} and are interested in using \systemname{} for their text-based analysis tasks.
This work has the potential to impact a number of data-driven fields that emphasize annotation quality and, in particular, benefit multidisciplinary areas that deal with critical problems such as maintaining \edit{the} vital infrastructure and ensuring community resilience.

Our main contributions can be summarized as follows:
\begin{itemize}[leftmargin=*, noitemsep, topsep=0px]

\item We propose a human-in-the-loop workflow that supports domain practitioners to efficiently conduct validation and relabeling tasks for large-scale technical text annotations from weak supervision.

\item We encapsulate this workflow as a web-based interactive notebook with a visual analytic interface that facilitates the identification of annotation errors and relabeling for different scales of data.

\item We collect insights from domain experts in different application domains and observe various preferences corresponding to their backgrounds, which shed light on directions for improving \systemname{} and cater to the needs of diverse domain practitioners.

\end{itemize}

\section{Related Work}
\label{sec:relwork}


\subsection{Technical Language Processing}
\label{subsec:TLP}



Process monitoring, diagnostics, and prognostics have gained prevalence with the increased emphasis on smart manufacturing, and reduced machine downtime. 
This trend---coupled with lower cost, more accessible
sensors and data storage solutions---has increased the volume of maintenance data~\cite{Brundage2019we}.
Despite the potential benefits, companies frequently struggle to adopt advanced manufacturing technologies due to cost of and lack of technical expertise in data analysis~\cite{Jin2016present}.
Simple yet powerful solutions for data analysis are necessary to aid manufacturers improve their practices.
There has been an increasing focus on sensor data and predictive maintenance using AI techniques~\cite{susto2014machine, carvalho2019systematic}.
However, these works often neglect a large part of maintenance data: natural language contained short-text maintenance logs, which leads organizations to turn to NLP.

\edit{
Technical text, however, poses challenges to commonly used NLP methods. Technical fields are often low-resource settings from an NLP perspective; they lack available resources such as annotated data and algorithms appropriate for specific analyses~\cite{Dima2022-ds}.
Transfer learning is the traditional strategy for addressing low-resource domains in machine learning~\cite{Dima2021-ba}. 
Models that were generated from annotated data from resource-rich domains are adapted for the low-resource domain.
Transfer learning approaches often assume limited differences between two different domains. 
But the technical text that appears in industrial information systems deviates considerably from ``standard'' English~\cite{Dima2021-ba}, full of expressions like ``1 W Mech Insp Ball Mill BM001'' and ''DSHT Cons Thkner rplace bed press''.

The lexical, grammatical, and terminological differences between ``standard'' English and industrial technical text have spawned bespoke domain-specific NLP adaptations that are largely outside of mainstream NLP~\cite{Dima2021-ba}.
}
TLP is a human-in-the-loop, iterative approach that addresses perceived shortcomings of applying standard NLP (natural language processing) to technical text data~\cite{Dima2021-ba}.
Originating with manufacturing maintenance, it is an adaptation of NLP that focuses on the technical text communicated within specialized domains.
TLP emphasizes the practical importance of semantic information and extends its system boundaries beyond algorithms and pipelines to include human input and community resources~\cite{brundage2021technical}.
The short-text from maintenance work orders (MWOs) are important analysis corpora for TLP~\cite{Lukens2019best, Sexton2017hybrid}. 
They record in detail the maintenance history of equipment and collectively capture vital information about inspections, diagnoses, and corrective actions~\cite{brundage2021technical}.
Annotation methods for MWOs
have been the subject of recent research in TLP.
Tools, such as Nestor\footnote{\href{https://nist.gov/services-resources/software/nestor}{https://nist.gov/services-resources/software/nestor}} have been developed to support the manual injection of critical real-world knowledge by allowing for the annotation of the MWO text descriptions via tagging to facilitate automated categorization and analyses.
Machine learning systems can then use these tags as a signal to help ensure correct outcomes~\cite{Dima2021-ba}.

\subsection{Large-Scale Text Annotation}
\label{subsec: weaksup}

The exponential growth of text data has made the current manual text annotation approaches\edit{, e.g., crowd-sourcing~\cite{jeff2006crowdsource},} deficient in meeting the pressing demands for high-quality large-scale annotations~\cite{young2018recent,lison2021skweak}.
\edit{As an alternative, researchers have developed the weak supervision techniques~\cite{ratner2017snorkel,fu2020fast,lison2021skweak} that leverage human-defined \second{labeling functions (LFs)}~\cite{ratner2017snorkel}, small labeled datasets~\cite{varma2018snuba}, or existing text paradigms with multi-type metadata~\cite{mekala2020meta} for more efficient text annotating.}
However, most of these approaches trade off labeling speed or cost with annotation quality~\cite{ratner2017snorkel, varma2018snuba, mekala2020meta, lison2021skweak}, and the generated labels are mainly evaluated by numerical performance matrices, such as accuracies~\cite{ratner2017snorkel}, F1 scores~\cite{varma2018snuba}.
Without human review, it is uncertain whether such annotations are of sufficient quality for real-world applications.
\edit{
In light of the deficiencies of manual and automatic text annotation approaches, a series of semi-automatic text annotation frameworks have been proposed, allowing humans to annotate large-scale text data with the help of automatic modules, which can be coordinated labeling modules~\cite{zhang2022onelabeler} or deep learning techniques such as attention model~\cite{choi2019aila}, human-validated labeling functions~\cite{evensen2020ruler,rietz2021cody}, and transductive semi-supervised learning~\cite{desmond2021increasing}.
However, the annotation quality of such frameworks still lacks human validation---they either only verify quantitative performance matrices~\cite{zhang2022onelabeler}, or sample a small subset for humans to inspect results~\cite{evensen2020ruler,desmond2021increasing}.
Although there are a few works for improving the annotation quality~\cite{liu2018interactive,bauerle2020classifier,tang2021videomoderator}, they are mainly designed for image or video data and hence not directly applicable to technical language datasets.
Given the importance of high-quality annotations~\cite{matsuda2009assessment, dumitrache2015achieving}, a human-centered tool is needed to support the validation of large-scale text annotations.
}

\subsection{Technical Text Visualization}
\label{subsec:TLP-vis}
In the past decade, the idea of applying visualization and visual analytics to technical text analysis has been broadly embraced.
Manufacturing enterprises are becoming aware of the value of maintenance records they collect and are supporting visualization research~\cite{alsallakh2016powerset, chen2017sequence, guo2017eventthread, chandrasegaran2020using}.
Academic researchers have developed visual analytical strategies for maintenance records ~\cite{zhang2021visual} and error logs~\cite{shilpika2019mela, Kesavan2020visual, Li2019visual, Nguyen2019visualizing}.
In particular, La VALSE~\cite{Guo2018valse} and MELA~\cite{shilpika2019mela} are scalable visualization tools with multiple visualization interfaces incorporating different logs for interactive event analysis.
ViBR~\cite{Chan2019vibr} provides a visual summary of
large bipartite relationships by via minimum description lengths and is used for vehicle fault diagnostics.
However, existing solutions have prerequisites on either the text format or the quality of the labels.
Some assume that there exists a well-defined set of labels~\cite{hamooni2016logmine} to train a classification modelfor the annotation task or assume a trivial effort to define these labels in the pre-processing stage ~\cite{di2017logaider, zheng2009system} when they are not available as input.
Others expect that the text can be generated from grammar or rules so that the labels can be derived from clustering~\cite{Guo2018valse}.

In this work, we address inconsistent technical text created by human maintainers that contains domain jargon and labels of unknown reliability. Unlike other approaches, we do not have prerequisite text formats nor do we make assumptions about the labels or their quality.
We also do not rely on the text's grammatical structure

\begin{figure*}[t]
  \centering
  \includegraphics[width=0.8\textwidth]{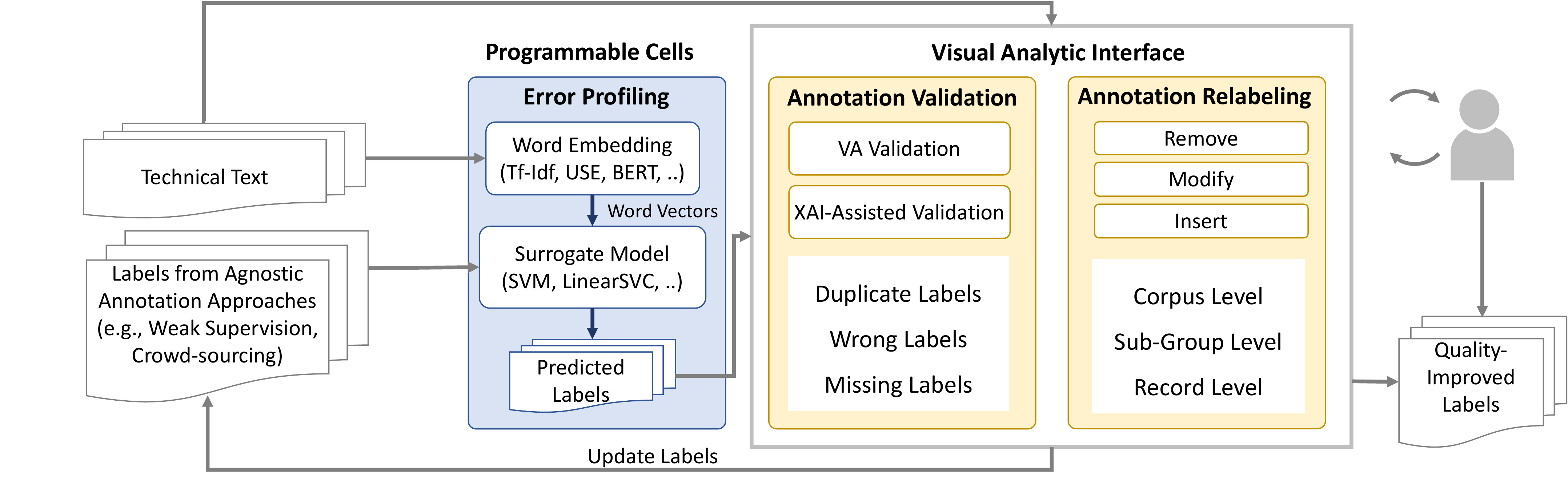}
  \vspace{-3mm}
  \caption{
    The \systemname{} workflow consists of three phases: Error Profiling, Annotation Validation, and Annotation Relabeling. It is implemented as a web-based interactive notebook which takes technical text and the corresponding labels as input. With the actionable insights provided by the surrogate model, XAI method, and visualization, users can identify three error types and improve annotation quality at three different data scales. 
  }
  \vspace{-3mm}
  \label{fig:workflow}
\end{figure*}

\section{Validation and Relabeling for Text Annotations}
\label{sec:problem_definition}
In this section, we define the problem we wish to solve and clarify our assumptions.
We also derive the annotation error types and design requirements based on \edit{the industrial experience of two of the authors} and an exploration of a machine maintenance log dataset.

\subsection{Problem Definition}
\label{subsec:definitions}

Like software development, TLP often requires machine-assistance in the validation of \textit{large}, \textit{complex}, and \textit{context-specific} sets of text.
To fulfill this need, we aim to facilitate the validation of  annotations and the correction of labelling errors by designing visual analytic techniques for domain practitioners.

Our target users are domain practitioners and data analysts in need of assessing and improving the annotations for large-scale technical text data.
We expect they have the analytical skills to interpret the model performance metrics and interact with \systemname{}.
We also make two assumptions about the technical log text and their labels: 

\begin{itemize}[leftmargin=*, noitemsep,topsep=0px]
    \item \label{assmpt:labels} 
    There exists a finite set of labels ${L} = \{l_1, l_2, ..., l_n\}$. 
    And the mapping from each record $s_i$ to the labels is defined by $l_i=f(s_i)$, where $l_i \subseteq L$. 
    In the context of this paper, $f(s_i)$ is agnostic and the quality of ${L}$ requires expert verification.
    \item \label{assmpt:category} 
    There exists a finite set of label categories ${C} = \{c_1, c_2, ..., c_n\}$. Each label $l$ in ${L}$ belongs to one label category in ${C}$. 
    Note that the ``label category'' in our context is a higher-level taxonomy of labels.
    For instance, label ``\emph{air-conditioner}'' belongs to \dataset{Item} and ``\emph{too\_hot}'' belongs to \dataset{Problem} in Sec.~\ref{subsec:use_case_HVAC}. 
\end{itemize}

\edit{
Given there is no formalized taxonomy of labeling errors in the TLP domain, we target three dominant types of annotation errors distilled from two coauthors' long-term industrial experience:}
\begin{itemize}[leftmargin=*, noitemsep,topsep=0px]
\item \label{error: duplicate} \textbf{Duplicate Labels} share duplicated words (e.g., ``\emph{temperature}'' and ``\emph{room\_temperature}'') and/or express semantic meanings (e.g., ``\emph{too\_cold\_building}'' and ``\emph{temperature\_too\_cold}'').
\item \label{error: wrong} \textbf{Wrong Labels} involve labels with conflicting meanings (e.g., one record is labeled with ``\emph{too\_cold}'' and ``\emph{too\_hot}'' simultaneously). \edit{It can also refer to} an unreasonable label name (e.g., ``\emph{building\_building}'') \edit{where the label refers to a non-existent class.}
\item \label{error: missing} \textbf{Missing Labels} refer to labels that \edit{should} be assigned to technical records but are absent (see examples in Sec.~\ref{subsec:use_case_HVAC}).
Missing labels are relatively hard to detect if there are already other labels assigned to the record.
\end{itemize}
The output of our workflow is a set of labels with improved quality. 


\subsection{Design requirements}
\label{subsec:design_requirements}
\edit{After much discussion on a weekly base, our team, which included two TLP domain experts, agreed to four design requirements for \systemname{} to address the problem defined in Sec.~\ref{subsec:definitions}}:


\begin{enumerate}[start=1,label={\bfseries R\arabic*},leftmargin=*, noitemsep,topsep=0px]

    \item \label{req:overview} \textbf{Label Overview}: 
     As the first step of label debugging, \systemname{} needs to provide users with a summary overview of all technical text and labels. 
     The visual interface needs to present label distribution in the finite label set ${L}$ and illustrate their categories $c$ (if available) intuitively.
     The visual interface also needs present the currently assigned labels of one or multiple record(s) $s_i$ in different levels of detail and from different perspectives per user demand to support intensive context comprehension.

    \item \label{req:detection} \textbf{\edit{Label Quality Screening}}: 
    \systemname{} \edit{need to support an efficient evaluation}
    of the quality of existing annotations.
    In particular, the visual interface needs to allow users to quickly locate labels that potentially fall into the three types of errors (see Sec.~\ref{subsec:definitions}).
    After that, it should help users confirm the error by providing sufficient context information about the related labels and explaining how they were assigned to specific records.
    
    \item \label{req:relabel} \textbf{\edit{Interactive Relabeling Support}}:
    Once the errors are identified and confirmed, \systemname{} needs to interactively collect the user's relabeling suggestions and apply them to specific scales of the dataset per user request.
    In particular, users should be able to make suggestions to remove or modify an existing label or insert new labels according to their best judgment.
    After that, such modifications should be applied to entire corpus, a sub-group, or an individual record per user demand.
    
    \item \label{req:flexibillity} \textbf{Accessibility and Flexibility}:
    \systemname{} should be accessible to domain practitioners of varying backgrounds.
    On the one hand,
    the basic functionalities of the visual interface should be intuitive enough for users without a computing background during the validation and relabeling tasks.
    On the other hand,
    \systemname{} should provide users with in-depth information on demand and the flexibility to adjust the data processing or model training settings so that the analysis process also satisfies experts with more computing experience and special analysis needs.

\end{enumerate}






\edit{
\subsection{Datasets}

We involve two TLP datasets, \textbf{HVAC} and \textbf{NLU}, in this paper:

\textbf{HVAC} is \second{an internal dataset from our industrial collaborators} with over $21,000$ pieces of maintenance records from an HVAC system.
Each record contains two text fields: \dataset{LONG\_DESCRIPTION} and  \dataset{DESCRIPTION}.
\dataset{LONG\_DESCRIPTION} describes the detailed maintenance information, including the problem, the solution, the maintainer, the corresponding machine, etc., while \dataset{DESCRIPTION} is a concise version, which is often a sentence or a set of keywords.
There are also eight categories of labels available for each record, including \dataset{P} (Problem), \dataset{S} (Solution), \dataset{I} (Item), \dataset{PI} (Problem Item), \dataset{SI} (Solution Item), \dataset{X} (Irrelevant), \dataset{U} (Unknown), and \dataset{NA}.
For example, the category \dataset{P} includes labels such as ``\emph{too\_hot}'', ``\emph{leak}'', and the category \dataset{SI} includes labels such as ``\emph{adjust thermostat}'', ``\emph{replace valve}'', etc.
These labels were produced by a weak supervision method, and their quality remains agnostic.

\textbf{NLU}~\cite{braun2017evaluating} contains over $25,000$ human-robot interaction records and the corresponding labels, collected from a voice AI agent serving in an intelligent home system.
Each record includes three text fields: \dataset{question} is a pre-designed human-robot interaction question; \dataset{answer} and \dataset{answer\_normalized} contain the original and normalized user answers, respectively.
There are three categories of labels, including \dataset{scenario}, \dataset{intent}, and \dataset{suggested\_entities}.
For example, the category \dataset{scenario} includes labels such as ``\emph{weather}'', ``\emph{music}'', and the category \dataset{intent} includes labels such as ``\emph{request}'', ``\emph{send email}'', etc.
These labels were generated from a crowd-sourcing platform, and their quality requires validation as well.
}

\section{Methodology}
\label{sec:methodology}


\subsection{Workflow}
\label{subsec:workflow}

We designed the \systemname{} workflow as an iterative framework with three major phases: (1) Error Profiling, (2) Annotation Validation, and (3) Annotation Relabeling.
A regular analysis process starts from the \textbf{Error Profiling} phase, in which we train a surrogate model with the technical text and their existing labels to approximate the prior annotation process.
Then, users can conduct the first round of \textbf{Annotation Validation} through the integrated visual analytic interface, where multiple coordinated views are provided to assist an efficient investigation of labels (\ref{req:overview}) and detection of three types of errors.
After that, users can move on to the \textbf{Annotation Relabeling} phase and relabel the identified results at three different levels: corpus level, sub-group level, and record level (\ref{req:relabel}).
A more detailed description of our visual and interactive support on these three levels is provided in Sec.~\ref{sec:visualization}.
After the first pass of the three phases, users can iterate between \textbf{Annotation Validation} and \textbf{Annotation Relabeling} for multiple rounds till the annotation quality converges with their standard of satisfaction.
It is also worth mentioning that \systemname{} simplifies the input and output of phase(1) so that users only need to make minor hyperparameter adjustments to execute different use cases with various analysis purposes (\ref{req:flexibillity}).

 


\subsection{Surrogate Model for Error Profiling}
\label{subsec:auto-labeling}


In the first phase of the \systemname{} workflow, we train a surrogate ML model~\cite{cozad2014learning} to approximate the generation process of existing labels in the dataset.
\edit{To ensure that the surrogate model can achieve satisfactory performance and reflect potential annotation issues, we tuned the model architecture with the interactive notebook to fit the specific dataset.
Then, by visualizing the model's intermediate results (e.g., prediction probability~\cite{jordan2010program}) in the second and third workflow phases, we help users uncover potential annotation flaws.}
In this way, users start their label validation from those suspicious labels related to unusual model behaviors (\ref{req:detection})
and locate a group of potential labeling errors for inspection.
After addressing these labels, users can retrain the surrogate model with the better-quality dataset to obtain a reusable model incorporating domain knowledge from human experts and save it for future annotation tasks.

\edit{The error profiling phase of \systemname{} require the surrogate model to be: (1) lightweight, so that the model tuning is time-effective; (2) accurate in producing similar results to existing annotations.
For (1), we utilize lightweight and time-effictive word-embedding and ML methods to process text data and train the annotation classifier.
For instance, to process the input technical text data, we adopt computationally efficient and widely-used word embedding techniques, including TF-IDF (term frequency-inverse document frequency)~\cite{enwiki:1071253989} and truncatedSVD (Singular value decomposition)~\cite{golub1971singular}, to encode the original text into real-valued vectors.
For (2), we iterate multiple processes with different model training settings, and audit quantitative performance matrices in the validation split until reaching the best result.
Thus, we ensure the surrogate model achieves satisfactory performance, i.e., the alignment between the predicted labels and the existing annotations is reasonable for the surrogate model to simulate the annotating process and provide hints for users.}
Specifically, the average hamming loss is $0.02$, the micro f1 score is $0.8044$, and the average macro f1 is $0.6703$, where smaller hamming loss, and larger micro \& macro f1 \edit{scores} indicate better performance.
Besides, the predicted probabilities $predProba$ of the fitted LinearSVC $fittedModel$ is obtained to act as the clue of finding suspicious labels.

\begin{figure*}[t]
  \centering
  \includegraphics[width=\linewidth]{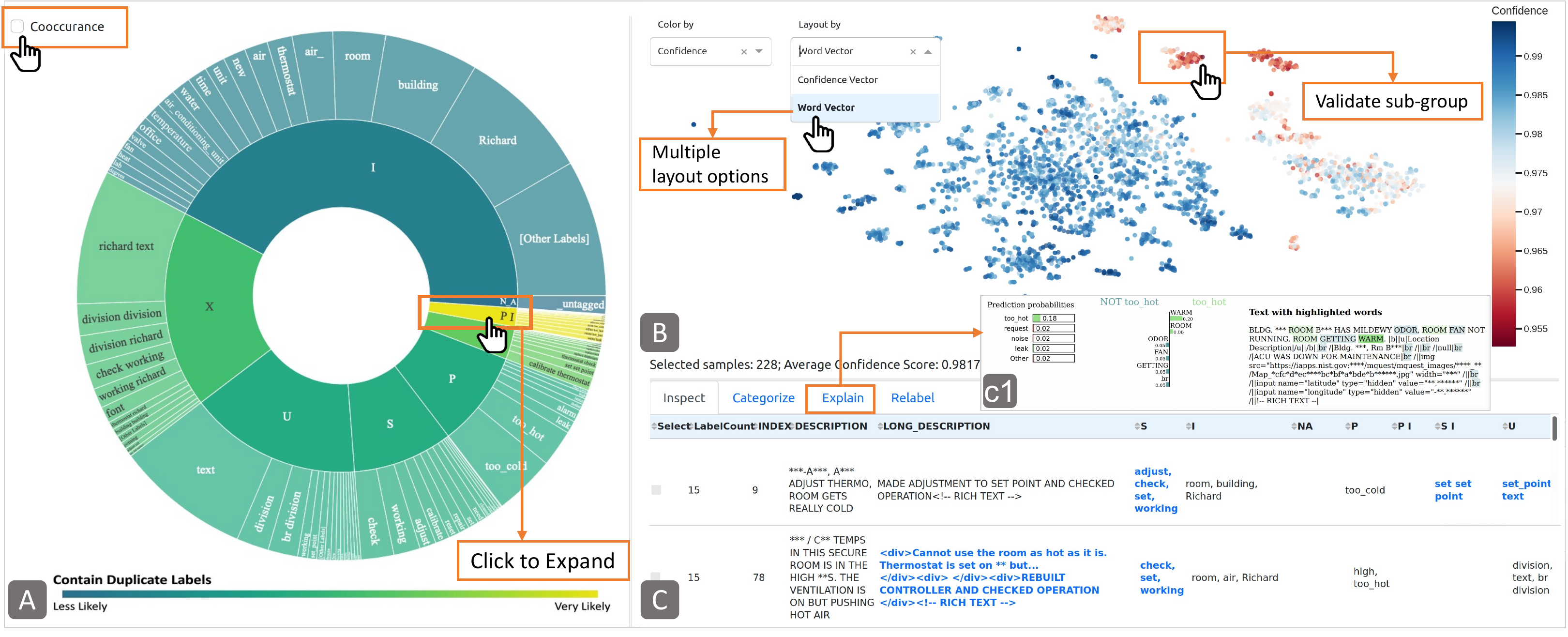}
  \caption{The \systemname{} interface with the HVAC dataset. (A) \textbf{\viewA{}} visualizes the label and category hierarchy relationships; each category can be expanded to present label co-occurrences (see Fig.~\ref{fig:HVAC1} (A)). \edit{(B)} \textbf{\viewB{}} presents the record distribution oto support sub-group validation, layout by model confidence vectors or input word vectors. The color represents ``model prediction confidence'' or ``record info density''. (C) \textbf{\viewC{}} includes multiple tabs, ``Inspect'' for record and label inspection, ``Categorize'' for category-based investigation, ``Explain'' for model behavior interpretation (c1), and ``Relabel'' for relabeling operations.
  }
  \vspace{-3mm}
  \label{fig:interface}
\end{figure*}

\subsection{Model Behavior Explanation}
\label{subsec:multi-level}

 \systemname{} utilizes one of the state-of-the-art eXplainable Artificial Intelligence (XAI) techniques--LIME (Local Interpretable Model-Agnostic Explanations)~\cite{ribeiro2016should}--to support the annotation validation.
For each of the constructed surrogate models, LIME performs perturbation-based analysis over a given text record and presents the explanation by highlighting the rationale behind the model's prediction.
It exposes the weakness of the model and the pitfalls of the input technical text and thus could help users more accurately inspect a potential annotation error and make a reasonable relabeling decision.
The LIME explanation is integrated into the ``Explain'' tab of \viewB{} \systemname{} and is triggered when users select a record from the ``Categorize'' tab for further inspection (more details in Sec.~\ref{sec:visualization} and examples Sec.~\ref{sec:scenario}).

\subsection{Implementation}
\label{subsec:implementation}
To maximize its accessibility for sharing and flexibility for customization (\textbf{R4}), we implement the \systemname{} workflow as a computational notebook.
We use multiple Python data analysis libraries including Pandas~\cite{reback2020pandas}, Numpy~\cite{harris2020array}, and Joblib~\cite{joblib2020} for data processing and intermediate metrics analysis.
In addition, the word embedding techniques (TF-IDF and truncated SVD) and ML methods (LinearSVC) discussed in Sec.~\ref{subsec:auto-labeling} are implemented with scikit-learn~\cite{scikit-learn}, and the LIME technique is with LimeTextExplainer~\cite{ribeiro2016should}.
To ensure smooth integration and faster rendering speed, we embed the visual interface (Fig.~\ref{fig:interface}) in the computational notebook with Plotly's JavaScript Graphing Library and Plotly Dash.
And we use the t-distributed stochastic neighbor embedding (t-SNE) algorithm for dimensionality reduction when visualizing the high-dimensional word vectors and confidence vectors for the \viewB{}.
We also deliberately separate the functions in the notebook so that users can easily plug in any word embedding and dimensionality reduction algorithms for their specific analysis needs with minor programming.

\begin{figure*}[t]
  \centering
  \includegraphics[width=0.9\textwidth]{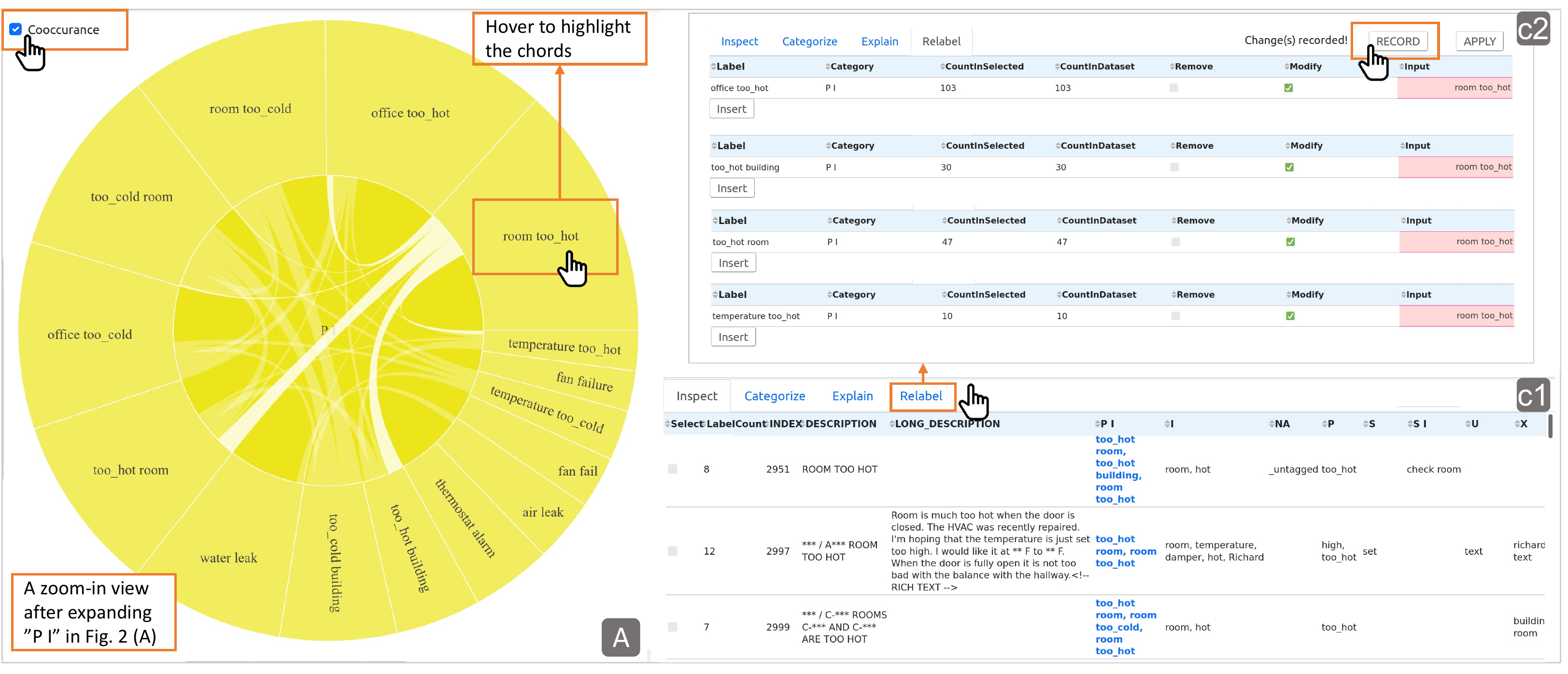}
  \caption{
  Duplicate label validation with the HVAC dataset (Sec.~\ref{subsec:use_case_HVAC}). The chord diagram (A) shows that labels ``\emph{office too\_hot}'', ``\emph{room too\_hot}'', 
  ``\emph{too\_hot building}'', and ``\emph{too\_hot room}'' co-occur frequently. Duplication is confirmed via record details in (c1) and fixed in ``Relabel'' tab (c2).  }
  \vspace{-3mm}
  \label{fig:HVAC1}
\end{figure*}

\section{Visual Analytic Interface}
\label{sec:visualization}
To fulfill the design requirements in Sec.~\ref{subsec:design_requirements} and concretize the \systemname{} workflow in Sec.~\ref{subsec:workflow}, we design a visual analytic interface (see Fig.~\ref{fig:interface}) that contains three major components: (A) \viewA{}, (B) \viewB{}, and (C) \viewC{}.
In this section, we demonstrate how we can coordinate these views to locate the three types of errors 
and perform multi-level validation and relabeling on annotations.

\subsection{Annotation Validation}
\label{subsec:vis-validation}

With the coordination among different components of \systemname{}, users can efficiently validate the annotation quality and identify the three major types of error introduced in Sec.~\ref{subsec:definitions} (\textbf{R2}).

\subsubsection{Duplicate Label Detection}
\label{subsec:vis-duplicate}
\edit{To support duplicate label detection, we designed}
the \viewA{} (Fig.~\ref{fig:interface} (A)) \edit{and}
the ``Inspect'' tab of \viewC{}.
\edit{We choose the sunburst diagram for \viewA{} to provide an overview of the hierarchical relationship between labels and their category (\textbf{R1}), as well as the distribution of the labels across categories -- }the size of the label sectors at the outermost layer represents the number of records in the dataset assigned with the corresponding label.
\edit{We also encode the possibility of duplicate labels into the sectors colors to provide a priority recommendation for the user inspection.}
\edit{This duplication possibility is the average of a} ratio of co-occurrence number $Co(l_i,\ l_j)$ to the total appearance $Num(l_i)$ of each label $l_i$ in the category:

\begin{equation}
P_{duplication} = \frac{1}{n_{categ}}\sum_{i=1}^{n_{categ}}(\frac{1}{n_{occur}}\sum_{j=1}^{n_{occur}}\frac{Co(l_i,\ l_j)}{Num(l_i)})
\end{equation}


\edit{The sunburst diagram is expandable per user request.
And in the zoom-in view, we embed a chord diagram to illustrate the label co-occurrence in the same record, which is a strong indicator of duplicate labels (e.g., Fig.~\ref{fig:HVAC1} (A)).}
\edit{In this chord diagram,} the co-occurring labels are connected with white chords, \edit{with their thickness representing} the co-occurrence frequency.
For example, the chord between the label ``\emph{room too\_hot}'' and ``\emph{too\_hot room}'' is thicker than that between ``\emph{room too\_hot}'' and ``\emph{water leak}'', indicating a heavier co-occurrence pattern and potential duplication of the former pair.
A larger number of thicker chords also indicates a higher possibility of existing duplicated labels in this category, which corresponds to the brighter sector color described above.


Users can further inspect the context of any suspect labels by clicking on them and checking the updated data table under the ``Inspect'' tab in the \viewC{} (Fig.~\ref{fig:interface} (C)).
This data table presents all the records across the dataset assigned with the selected label.
In this way, users can efficiently locate and evaluate the correctness of potential problematic labels.


\subsubsection{Wrong Label Detection}
\label{subsec:vis-wrong}
Wrong labels can be detected with the \viewB{} (Fig.~\ref{fig:interface} (B)) in coordination with the ``Categorize'' and ``Explain'' tabs of \viewC{}.
\edit{To provide a two-dimensional (2D) overview for all records (\textbf{R1}), we apply the t-SNE~\cite{van2008visualizing} algorithm to project the customized record vectors onto the 2D space and visualize each of them as a dot.}
The customized record vector can be a ``word vector'' or a ``confidence vector''.
When the ``word vector'' is used for layout, the distance among the dots indicates the semantic closeness of descriptions in their corresponding records.
When the ``confidence vector'' is used for layout, the distance among dots indicates the model behavior towards similarity when predicting labels for the corresponding records.
We provide two options to color the record projections --- ``information density'' and ``confidence score''.
The ``information density'' is more useful in locating missing labels, so we will discuss it in Sec.~\ref{subsec:vis-missing}.
The ``confidence score'' is the mean value of all dimensions of the aforementioned ``confidence vector'', which could expose the records containing more labels predicted with low confidence, and thus provide hints to locate sub-groups that potentially contain labeling mistakes (Fig.~\ref{fig:interface} (B)).


Once a cluster with low confidence is identified and selected, a heatmap under the ``Categorize'' tab in \viewC{} (Fig.~\ref{fig:HVAC2} (c1)) will be triggered, where each row refers to a single record; each column represents one label category, and the color indicates the model's average confidence score.

\edit{To support deeper understanding of the model reasoning process, we provide LIME explanations under the ``Explain'' Tab (Fig.~\ref{fig:interface} (c1)).}
The explanation includes three parts:
the left bar chart visualizes the top five predicted labels and their prediction probabilities;
the middle bar chart visualizes the ``score of contribution'' of the input words to the top label;
\edit{the right side shows more context information and the original text record,}
where the positive and negative contributors are highlighted with different colors.
Combining these three kinds of information, 
\edit{we aim to help users verify whether the rationale behind the model's decision aligns with their knowledge (\textbf{R2}).}

\subsubsection{Missing Label Detection}
\label{subsec:vis-missing}
The detection of missing labels also involves the \viewB{} (Fig.~\ref{fig:interface} (B)) and ``\edit{Inspect}'' tabs of \viewC{}.
\edit{We designed the ``information density'' metric to highlight records more likely to have missing labels.}
This metric is determined by the ratio of label count and the input text length:

\begin{equation}
D_{Info} = log(\frac{Count(labels)}{WordCount(text)})
\end{equation}

Once the users locate and select a cluster of records with low ``information density'', the ``Inspect'' tab of \viewC{} will be updated for verification of the label missing issue.
It is also worth mentioning that higher ``information density'' can insufficiently indicate the existence of duplicate labels, but further verification is required with the process in Sec.~\ref{subsec:vis-duplicate}.

\subsection{Annotation Relabelling}
\label{subsec:vis-record}

After confirming a labeling error, users can improve the annotation quality of the dataset (\textbf{R3}) in three different data scales: corpus level, sub-group level, and record level:
\begin{itemize}[leftmargin=*, noitemsep, topsep=0px]

\item\textbf{Corpus level relabeling} updates the label across the entire technical text dataset.
It is achieved by clicking on a label from \viewA{} and use the ``Relabel'' tab of \viewC{} to ``remove'', ``modify'', or ``insert'' it.
For example, the label ``\emph{building\_building}'' is considered to be a ``wrong label'' error at the corpus level.
Users can select it from \viewA{} and remove it from all affected records.

\item\textbf{Sub-group Level relabeling} involves a sub-group of records within the dataset.
It is achieved by selecting a sub-group of records with the lasso tool in \viewB{} and relabeling them with the ``Relabel'' tab.
The number of records in one sub-group can vary from a dozen to hundreds.

\item\textbf{Record level relabeling} updates individual record(s) associated with a specific label error.
For example, a user looks over the records through the ``Inspect'' tab and notice two records missing the label ``alarm''.
They can select these two record(s) with the checkbox and relabel them under the updated ``Relabel'' tab.
These relabeling operations are only applied to the selected records.
\end{itemize}
\edit{To eliminate waiting for dataset updates and projection re-rendering, the visual interface is only re-rendered when the user request to apply the changes.
To achieve this, we record the relabeling operations as a history list and sequentially apply them to the database per request.}

\section{Use Case Scenarios}
\label{sec:scenario}

This section describes two use case scenarios where \systemname{} assists domain experts in validating the quality of technical text annotations and conducting efficient relabeling. 

\subsection{Case 1: Maintenance Management for HVAC System}
\label{subsec:use_case_HVAC}


This use case involves Amy, a maintenance manager who monitors machine maintenance records to track maintenance-related issues and to plan for future maintenance resources (e.g. budgets, maintainers, etc.).
With the HVAC dataset, we demonstrate how she uses \systemname{} to validate the label quality of MWOs and make maintenance management based on the more accurate labels.


\edit{
After finishing data processing and surrogate model training in the programmable cells of \systemname{}, Amy starts validating labels through the interface.}
Because the frequency of similar labels reflects the prevalence of a maintenance issue and influences decision priorities and budgets, Amy chooses to screen duplicate labels at first.
She notices from the \viewA{} (Fig.~\ref{fig:interface} (A)) that the category \dataset{PI} is the most likely to contain duplicate labels, so she expands it to check for label co-occurrance (Fig.~\ref{fig:HVAC1} (A)).
As indicated by the chord thickness, ``\emph{office too\_hot}'', ``\emph{room too\_hot}'', ``\emph{temperature too\_hot}'', ``\emph{too\_hot building}'', and ``\emph{too\_hot room}'' co-occur very frequently.
Because these labels have similar semantic meanings, Amy further inspects their context in the ``Inspect'' table (Fig.~\ref{fig:HVAC1} (c1)) and confirms that they are duplicates.
Such duplication will overemphasize air-conditioner-related ``\emph{too hot}'' issues and may cause excessive allocation of maintenance resources.
Amy removes the redundant labels and unifies the rest with ``\emph{room too\_hot}'' with the ``Relabel'' (Fig.~\ref{fig:HVAC1} (c2)).
Now that Amy has created more accurate and consistent labels for temperature-related problems, she counts their frequency, evaluates the problem severity, and decides to arrange for regular examination of all air-conditioners across the company.

\begin{figure*}[t]
  \centering
  \includegraphics[width=.9\textwidth]{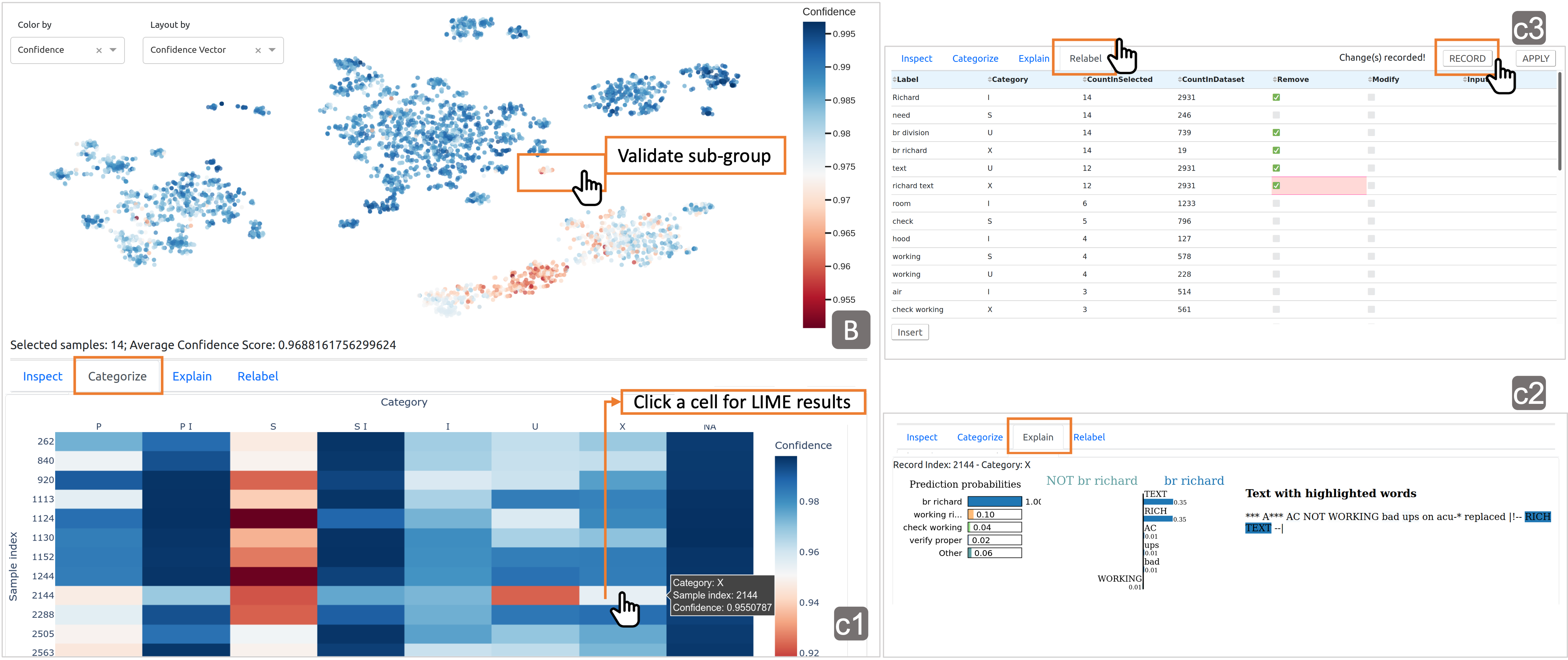}
  \caption{
  Wrong label detection in HVAC dataset using \systemname{} (Sec.~\ref{subsec:use_case_HVAC}). Users can select a sub-group with lower confidence in (B) and inspect model confidence of each category in (c1). Then they can click on cells in (c1) to activate LIME explanation in (c2) for model behavior interpretation. After confirming the error, users can remove the wrong labels with (c3).
  }
  \vspace{-3mm}
  \label{fig:HVAC2}
\end{figure*}

Amy moves on to the \viewB{} (Fig.~\ref{fig:interface} (B)) to screen for wrong labels which are another type of error that can mislead maintenance planning.
Amy notices that there are several clusters with lower confidence scores (Fig.~\ref{fig:HVAC2} (B)), indicating that the surrogate model performed worse and might predict the wrong labels for the corresponding records.
After selecting one, she learns from the updated ``Inspect'' tab that the labels ``\emph{Richard}'' in category \dataset{I} and ``\emph{br richard}'' in category \dataset{X} appear in many records.
Amy browses the context to check if this is related to a maintainer's name or is an improper description but finds the phrase ``br richard" doesn't appear to be semantically relevant to either \dataset{DESCRIPTION} or \dataset{LONG\_DESCRIPTION}.
Seeing this issue also appears in many other clusters, Amy decides to further investigate its cause with the ``Categorize'' and ``Explain'' functions.
Under the ``Categorize' tab, she clicks on the cell indicating lowest prediction confidence in category \dataset{I} and \dataset{X} (Fig.~\ref{fig:HVAC2} (c1)) to trigger a LIME explanation.
As shown in the left bar chart in Fig.~\ref{fig:HVAC2} (c2), the model's top prediction for the selected record is the label ``\emph{br richard}''.
However, the middle and right part shows that most positive contributors to this prediction are HTML metadata such as ``TEXT'' and ``RICH''.
Amy realizes that the wrongly-predicted label ``\emph{br richard}'' might originate from the presensce of HTML tags.
After seeing similar LIME explanations for more records in this cluster, her hypothesis is confirmed -- the model referred to HTML tags to predict the wrong labels.
Amy removes these wrong labels (Fig.~\ref{fig:HVAC2} (c3)) and decides to conduct a thorough cleaning of the dataset later to remove HTML tags.

Amy also checks the info density and does not find any severe label missing issues.
Then she reloads the updated labels into \systemname{} and confirms that the new annotations are satisfactory.
Finally, Amy executes the code cells of the computational notebook (Sec.~\ref{subsec:implementation}) to re-train the surrogate model with the relabeled dataset.
In this way, she preserves her domain knowledge in the model that can be used to annotate any future maintenance records.


\subsection {Case 2: Data Cleansing for NLU Model Training}
\label{subsec:use_case_NLU}
In this section, we demonstrate how \systemname{} can help modify the annotations of the training data for a specific application scenario.
This use case involves Steven, a data engineer working in a company that provides conversational agent (CA) services.
Steven coordinates the large-scale crowd-sourcing process to provide high-quality training data for the natural language understanding (NLU) model \cite{braun2017evaluating} embedded in the CAs, which summarizes the semantic content of user utterances by mapping it to structured, abstract representations (labels) that support the decision making process.



\begin{figure*}[t]
  \centering
  \includegraphics[width=.9\textwidth]{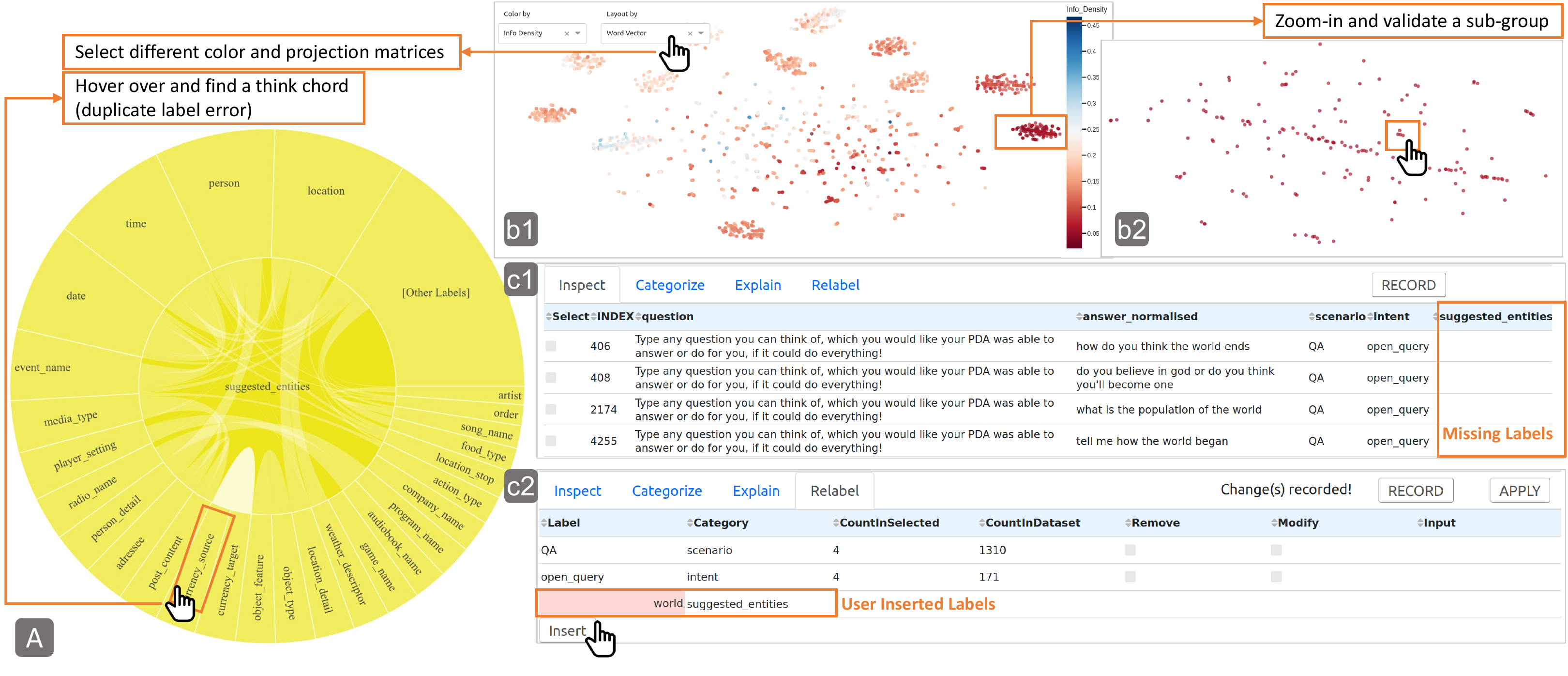}
  \caption{
       Finding duplicate and missing labels in NLU dataset using \systemname{} (Sec.~\ref{subsec:use_case_NLU}). The chord diagram in the \viewA{} (A) reveals the duplicate labels ``\emph{currency\_source}'' and ``\emph{currency\_target}''. The clusters with low information density in the \viewB{} (b1) highlight records with missing labels. 
        The ``Inspect'' view (c1) shows that the \dataset{suggested\_entities} for the selected records in (b2) are missing. User-suggested labels can be inserted through the ``Relabel'' tab (c2).
  }
  \vspace{-3mm}
  \label{fig:NLU2}
\end{figure*}

Steven uses \systemname{} to validate and debug the crowd-sourced annotation results before delivering the dataset for the downstream machine learning tasks.
The NLU model requires all labels to be independent and accurate so that the voice AI agent can exclusively query them in the search engine and provide correct answers to the users.
To remove those dependent labels with semantic duplication, so Steven starts by looking for them via the \viewA{}.
According to the coloring of the categories, the category \dataset{suggested\_entities} is most likely to include duplicate, so Steven expands this category to inspect the label co-occurrences.
The thick chord indicates that the labels ``\emph{currency\_source}'' and ``\emph{currency\_target}'' heavily overlap (Fig.~\ref{fig:NLU2} (A)).
Steven inspects the detailed context of the corresponding records with the \viewC{} and finds that most of these records are related to currency exchange questions.
Although the two labels ``\emph{currency\_source}'' and ``\emph{currency\_target}'' appear reasonable, Steven still decides to merge them into the single label ``\emph{currency\_source\_and\_target}''
to facilitate the down-streaming task.

When using the ``Inspect'' tab to investigate the duplicate label issues above, Steven notices that missing labels are a common with this dataset.
He moves on to the \viewB{} to facilitate find more missing labels.
He chooses the options of ``Color by Info Density'' and ``Layout by Word Vector'' to highlight clusters with similar semantic meanings and lower info density (Fig.~\ref{fig:NLU2} (b1)).
Then he uses the lasso tool to select the most notable cluster and observes from the \viewC{} (Fig.~\ref{fig:NLU2} (c1)) that all the records in this cluster were labeled as ``\emph{QA}'' under the category \dataset{scenarios} and ``\emph{object\_query}'' under the category \dataset{intent}, but not assigned with any labels under the category \dataset{suggested\_entities}.
With a second look at the input \dataset{questions}, \dataset{answers}, along with the model's reasoning process from the ``Explain'' tab, Steven figures out the cause --- the model only captured the information from the records' shared \dataset{question} but ignored the \dataset{answers}.
To fill the missing labels, Steven zooms into this cluster in the \viewB{} (Fig.~\ref{fig:NLU2} (b2)) and uses the lasso tool to select each sub-clusters with similar semantic meaning.
After selecting one sub-cluster (Fig.~\ref{fig:NLU2} (b2)), Steven notices similar semantic meanings of the selected records -- for \dataset{answer\_normalized}, those records have sentences such as ``\emph{how do you think the world ends}'', ``\emph{tell me how the world begin}'' and  ``\emph{do you believe in god}'', etc.
Considering these questions were asked by their CA users, Steven believes ``\emph{apocalypticism}'' or ``\emph{philosophy}'' would be proper labels for the category \dataset{suggested\_entities}.
He inserts them into the selected sub-group of records with the help of the ``Relabel'' tab.
Steven conducts the same operation to the few other low-info-density clusters, and fix the label missing issues accros the entire dataset. 

Finally, Steven applies his relabeling operations to the dataset and updates the interface.
After confirming the quality of the annotations, he delivers the dataset to the machine learning engineers for the downstream training tasks.
With \systemname{}, Steven optimizes the annotation quality from crowd-sourcing results and avoids the potential flaws that can bias the training of the voice AI agent.

\vspace{-1mm}
\section{Expert Reviews}
\label{sec:expert_reviews}
\systemname{} was developed with the participation of TLP domain experts (Sec.~\ref{subsec:design_requirements}) over the course of two years.
To evaluate the generalisability of our workflow and reveal insights from or practical value to domain practitioners, we invited another two TLP domain experts (E1 and E2) and two experts from other domains (E3 and E4) into our expert review studies.

\subsection{Expert Demographics}
All experts were experienced in data analysis and performed data annotation tasks in their daily work.
E1 and E2 was research engineers from the TLP community,
who were familiar with and had worked on the analysis of the HVAC dataset (Sec.~\ref{subsec:use_case_HVAC}) before the study for several years.
E3 was an economist and statistician who analyzed large-scale tabular datasets to gain insights into community resilience. 
E4 was a research social scientist who manually annotated large-scale datasets about risk perception and evacuation decision-making, and was in need of speeding up this process. 

\subsection{Tasks and Setup}
\label{subsec: task}
\edit{We conducted two pilot studies to simulate the remote setup, test the \systemname{} execution environment (online Colab via a local Jupyter Notebook) and adjust the content of the tutorial sessions.}
\edit{Then we finalized} a semi-structured, open-ended expert review in which each expert was asked to explore one of the two datasets described in Sec.~\ref{sec:scenario}.
Based on their domain of expertise and familiarity with the dataset, E1 and E2 used the HVAC Dataset while E3 and E4 used the NLU Dataset.
We shared the original dataset and its documentation\footnote{\href{https://github.com/xliuhw/NLU-Evaluation-Data}{https://github.com/xliuhw/NLU-Evaluation-Data}} with E3 and E4 before the study so that they could get familiar with it in advance.
Because the existing annotation in the NLU Dataset is relatively clean, we used an adapted version with two manually inserted errors for each error type in Sec.~\ref{subsec:definitions} in the study.

The study was conducted online via a video conferencing where the domain experts accessed \systemname{} from Google Colab Notebook\footnote{\href{https://colab.research.google.com/}{https://colab.research.google.com/}} via their personal computers. 
We shared the tutorial document with the experts no less than two days before the study.
The online study session started with a 25-min tutorial session that combined an introductory presentation, a live demonstration, and the mini-tasks.
An example mini-task for the HVAC dataset was ``Please use \systemname{} to find one pair of duplicated labels under the category `PI' (Problem Item), and then suggest how to modify it with the `Relabel' tab''. 
After the tutorial session, the experts were asked to freely explore their assigned dataset to validate and relabel the annotations (20-25 minutes).
During this process, the experts followed the think-aloud protocol to verbalize their thinking and suggestions.
Finally, the experts responded to a questionnaire with their demographic information and general feedback of \systemname{}.

\subsection{Observations}
In our study, all experts appreciated the value of \systemname{} for facilitating annotation refinement and expressed willingness to use it in their daily work, \edit{describing it as \emph{a very good tool} (E2) that was ``\emph{helpful at a high level of quickly and...pleasantly...identifying issues than just scrolling through a spreadsheet}'' (E4).}
Meanwhile, we observed that domain experts with differing backgrounds interacted distinctly with \systemname{} and sometimes provided divergent comments towards the same features.
We categorize the their behaviors and feedback during the exploratory and describe them below.


\textbf{Learning Curve.}
Based on their familiarity with the dataset and the tool, domain experts required differing times to overcome the learning curve (\textbf{R4}).
For instance, E1 and E2 had previously worked on the HVAC dataset with other annotation tools, so they spent less time grasping \systemname{} compared to the other two experts (E3, E4).
E2 expressed great interest in the methodology ``\emph{under the hood}'' and asked many technical questions to understand the underlying mechanism during the tutorial session.
Although E3 and E4 needed more hands-on instructions about using our tool, they were capable of replicating the moderator's operations and accomplishing the exploration task after the tutorial sessions.
They praised our tutorial session design, saying ``\emph{it helped very much...after (the moderator) demonstrated, it very easy to replicate}'' (E2).


\textbf{Functionality.}
In all four studies, the experts were able to efficiently evaluate the quality of the annotations (\ref{req:detection}) and successfully accomplish the relabelling (\ref{req:relabel}) by coordinating information from the three major views of \systemname{} (Fig.\ref{fig:interface}).
Moreover, E1, E2, and E4 successfully mastered the relatively complex ``Categorize'' and ``Explain'' functions and utilized them to understand the root cause of a potential wrong label.
We also received requests for more delicate annotation manipulations and more complicated information support from experts with shorter learning curves, such as modifying the name of a label category (E2) or showing the percentage value of the duplicated labels (E1, E2).
However, experts with longer learning curves requested simpler operations and more exploration guidance from the tool, such as simplified projection view (E3) or ``\emph{pop-up reminders...to remind people what these different tools are for in a really obvious way}'' (E4).
How to support more delicate label manipulation as well as ensure the accessibility of \systemname{} ((\ref{req:flexibillity})) is an inspiring topic that we will discuss in Sec.~\ref{sec:discussion}.



\textbf{Visualization.}
Interestingly, experts with different backgrounds and experience using (semi-) automatic annotation tools also showed different preferences towards our two major visualization components -- the \viewA{} and the \viewB{}.
Though all experts expressed their favor of the \viewA{}, saying they ``particularly like the chord diagram'' (E3) because it was ``\emph{very helpful}''(E1), ``\emph{intuitive enough}'' (E2) and they ``\emph{haven't seen labels presented in this way}" (E4), E2 mentioned ``\emph{the co-occurrence is less useful because I don't have enough flexibility to dive down into why there's that co-occurrence}.''
For the \viewB{}, experts knowing more about machine learning (E1 and E2) picked it up faster and appreciated its value in finding wrong and missing labels better.
``\emph{I think the projection view is super useful},'' said E2.
They ``\emph{would like to see even more options of projection spaces and be able to play around with those}.''
In contrast, E3 felt the same function was too complex and required ``\emph{a lot of playing around}.''
E4 didn't even get a chance to try out the different projection options because of the time constraint.

\textbf{Interaction.}
During the study, we received precious interaction improvement suggestions, including more cross-view coordination, operation history tracking, and typing suggestions.
E1, E2, and E4 suggested more flexible interactions, such as cross-view Boolean operations and subset highlighting. 
For example, when E2 inspected the label ``\emph{time},'' they mentioned that this label might involve different types of redundancy according to their prior knowledge about the dataset.
As a result, they requested Boolean operations between the \viewA{} and \viewB{} to sift those records of their specific need.
E3 and E4 suggested providing ways to keep track of the editing history, such as an undo function (E3),  a history list (E4), and some hints of what the user has just clicked (E4).
E2 suggested adding typing suggestions for relabel tab, such as auto-complete or alternative recommendation functions.
These suggestions reflected the experts' tacit knowledge gained from their long-term annotation practice and will direct us to a more accountable annotation tool in the next development iteration. 

\section{Discussion}
\label{sec:discussion}
The observations and feedback from the expert reviews indicated that \systemname{} provides a means for domain practitioners to validate and \edit{relabel} the technical text annotations ``\emph{quickly}'' and ``\emph{pleasantly}''.
They also suggest potential future work for our workflow and tool.
Below, we organize the lessons learned.

\textbf{Accessibility v.s. Functionality.}
We observed in our expert review that 
users with less machine learning and (semi-) automatic annotation tool experience
may go through longer learning curves with  \systemname{}.
They requested more exploration guidance or relabelling recommendations when using the tool, while the other group of users requested more complex functions, saying \systemname{} was ``\emph{good to find gross errors, but not for perfectionism}''(E2).
This is understandable because we required domain experts with diverse backgrounds to learn a relatively complex system within a limited time.
We plan to alleviate this problem by leveraging user modeling techniques~\cite{ha2022unified} to analyze the user behavior and guide them to start from different levels of complexity.
This way, it will also be safe to extend \systemname{} with more intricate functions, as recommended.

\textbf{Automation v.s. Human Trust.}
As computer science researchers, we tended to incorporate more automation \edit{in} \systemname{} during the development process, which was discouraged by our collaborators with technical text annotation backgrounds.
We observed that most of the data analysts tended to be ``\emph{over conservative}'' (E2) and had to closely check the raw text before ``\emph{starting to believe the systems is working}''(E1).
They also said that many cases were ambiguous, so they tended to examine more context before making relabelling decisions.
Because of this, \systemname{} currently still involves considerable manual work, as demonstrated in Sec.~\ref{sec:scenario}.
\systemname{} also only provides recommendations and explanations instead of one-step relabelling suggestions to supply users with a comfortable amount of information.
Indeed, there were no complaints about too much manual work during the expert review but we did receive praise that our tool helped ``\emph{focus their energy}''(E4).


\textbf{Application Domains.}
\systemname{} was originally designed to serve as a component of technical language processing, but it is generalizable to other annotation verification tasks.
The error profiling process can take any natural language descriptions and their labels as input and allow users to perform validation and relabelling via the interface.
If label categories are available, as was for our use cases (Sec.~\ref{subsec:definitions}), there will be two layers in the Sunburst diagram of \viewA{}.
Otherwise, the Sunburst diagram will devolve into a pie chart, with other \systemname{} functionality remaining unchanged. 

\vspace{-1mm}
\section{Conclusion}
\label{sec:conclusion}
We presented \systemname{}, a human-in-the-loop workflow that can help domain experts efficiently validate and improve the quality of multi-labeled technical text annotations.
\systemname{} utilizes a web-based interactive notebook to enable flexible data processing and model training, and integrates a visual analytic system to leverage human knowledge in annotation relabeling.
The interface coordinates different visual components for multi-type error detection (duplicate, missing, and wrong labels) in different dataset scopes (corpus level, sub-group level, and record level), and provides a human-centered solution targeting the quality enhancement for large-scale text annotations.
We demonstrate the usability of \systemname{} via two use case cases, and \edit{four} experts evaluated the effectiveness of our workflow through a study consisting of one-on-one qualitative evaluations.
We believe our work will encourage the design of visual analyticsfor other domain-driven problems and inspire future research efforts in creating higher-quality annotations for larger-scale text datasets.


\section*{NIST Disclaimer}
\vspace{-1mm}
\edit{
The use of any products described in this paper does not imply recommendation or endorsement by the National Institute of Standards and Technology, nor does it imply that products are necessarily the best available for the purpose.
}
\vspace{-1mm}

\section*{Acknowledgement}
\vspace{-1mm}
\edit{
We are grateful to the domain experts who volunteered for our expert reviews and the anonymous reviewers that helped improve this paper.
This research is sponsored in part by the Grant No. 70NANB21H170 from the U.S. Department of Commerce, National Institute of Standards and Technology and a gift from Bosch Research.
}

\vspace{-2mm}



\bibliographystyle{abbrv-doi}

\bibliography{main}

\end{document}


\maketitle

In this document, we provide supplementary materials for the \systemname{} paper. 


\section{Surrogate model} \label{sec:dataset}

In this section, we provide more details on the processing and training for the surrogate model.
Before training the model, we first process the input text annotations, including the labels $labelText$ and the list of label categories $categList$.
As shown in Algorithm~\ref{alg:aut_label}, we adopt one-hot-encoding to transfer $labelText$ into numerical labels with values $0$ and $1$.
According to the category that each label belongs to, we divide the numerical labels $labelVectorTotal$ into different label categories.
For each category, a LinearSVC~\cite{scikit-learn} is trained by treating $wordVector$ as the input and $labelVector$ as the ground truth.
In this way, we construct a surrogate model as a label estimator for each label category, including fitted TF-IDF, truncatedSVD, and LinearSVC, which takes text as input and outputs the labels.
Finally, Algorithm~\ref{alg:aut_label} returns $wordVector$, $modelDict$, and $predProba$ for the use in upcoming phases of \systemname{}.

\begin{algorithm}[bht]
\caption{Surrogate model for simulating label generation}
\label{alg:aut_label}
\begin{algorithmic}[1]
\REQUIRE ~~\\
Technical text data, $techText$;\\
Programmatically-created labels in text format, $labelText$;\\
Label categories as a list, $categList$;
\ENSURE ~~  \\
Word vector encoded from text, $wordVector$;\\
Surrogate models for each category, $modelDict$;\\
Predicted probabilities of the models, $predProba$;\\
\STATE{$vectorTemp$ $\gets$ TF-IDF($techText$, $stopwords$)}
\STATE{$wordVector$ $\gets$ truncatedSVD($vectorTemp$)}
\STATE{$labelVectorTotal$ $\gets$ OneHotEncoding($labelText$)}
\STATE{$labelVectorDict$ $\gets$ Categorize($labelVectorTotal$)}
\STATE{$modelDict$ $\gets$ $\{$ $\}$}
\STATE{$predProba$ $\gets$ $[$ $]$}
\FOR{$category$ \textbf{in} $categList$}
\STATE{$labelVector$ $\gets$ $labelVectorDict[category]$}
\STATE{$fittedModel$ $\gets$ LinearSVC($wordVector$, $labelVector$)}
\STATE{$modelP$ $\gets$ Pipeline(TF-IDF, truncatedSVD, $fittedModel$)}
\STATE{$modelDict[category]$ $\gets$ $modelP$}
\STATE{$curProba$ $\gets$ $fittedModel.pred\_proba(wordVector)$}
\STATE{$predProba$ $\gets$ $predProba.concat(curProba)$}
\ENDFOR
\STATE{\textbf{return} $wordVector$, $modelDict$, $predProba$}
\end{algorithmic}
\end{algorithm}

\section{Background of TLP Experts}
As mentioned in Sec. 3.2, we derived our design requirements based on frequent discussions with two of the technical language processing (TLP) experts, who are also the coauthors of this paper.
In this section, we report their background and experience working with technical text analysis experience.

The first expert is an mechanical engineer from industry and a co-founder of the Technical Language Processing Community of Interest \footnote{https://www.nist.gov/el/technical-language-processing-community-interest}.
They specialize in the use of text analysis and network science for human-centric knowledge management in technical/domain-heavy situations. 
Their other research interests include design optimization, network analysis, research operations, and human-systems-integration.
They had been working on topics related to technical text analysis for seven years.

The second expert is a computer scientist from the industry and also a co-founder of the Technical Language Processing Community of Interest.
They specialize in infrastructure or platforms for science-oriented analytics.
They conducted the manual cleaning of the HVAC dataset (Sec. 3.1) before this project.
They had been working on topics related to technical text processing for six years.

\bibliographystyle{abbrv-doi}
\bibliography{sections/00_references}